\documentclass[twocolumn,useAMS,usenatbib]{mnras}

\usepackage{subcaption}
\usepackage{color}
\usepackage{amsmath}
\usepackage{graphicx}
\usepackage{amssymb}
\usepackage{psfrag}

\usepackage{etoolbox}
\makeatletter
\patchcmd\@combinedblfloats{\box\@outputbox}{\unvbox\@outputbox}{}{%
   \errmessage{\noexpand\@combinedblfloats could not be patched}%
}%
 \makeatother

\newcommand{\be}{\begin{equation}}
\newcommand{\ee}{\end{equation}}

\title[Gravitational waves from mountains]{Gravitational waves from magnetically-induced thermal neutron star
mountains}
\author[E. L. Osborne and D. I. Jones]{E. L. Osborne${}^1$,
         D. I. Jones${}^1$\thanks{d.i.jones@soton.ac.uk}\\ \\
         ${}^1$Mathematical Sciences and STAG Research Centre, University of Southampton, Southampton SO17 1BJ, U.K.}

\begin{document}

\pagerange{\pageref{firstpage}--\pageref{lastpage}} \pubyear{0000}
\maketitle

\label{firstpage}

\begin{abstract}

Many low-mass X-ray binary (LMXB) systems are observed to contain rapidly spinning neutron stars.  The spin frequencies of these systems may be limited by the emission of gravitational waves.  This can happen if their mass distribution is sufficiently non-axisymmetric.  It has been suggested that such `mountains' may be created via temperature non-axisymmetries, but estimates of the likely level of temperature asymmetry have been lacking.   To remedy this, we examine a simple symmetry breaking mechanism, where an internal magnetic field perturbs the thermal conductivity tensor, making it  direction-dependent.   We find that the internal magnetic field strengths required to build mountains of the necessary size are very large, several orders of magnitude larger than the inferred external field strengths, pushing into the regime where our assumption of the magnetic field having a perturbative effect on the thermal conductivity breaks down.  We also examine how non-axisymmetric surface temperature profiles, as might be caused by magnetic funnelling of the accretion flow, lead to internal temperature asymmetries, but find that for realistic parameters the induced non-axisymmetries are very small.    We  conclude that, in the context of this work at least, very large internal magnetic fields are required to generate mountains of the necessary size.

\end{abstract}


\section{Introduction} \label{sect:intro}

Gravitational wave astronomy is now a reality, with the detection of $10$ binary black hole coalescences, and one binary neutron star coalescence by the Advance LIGO--Virgo network during its first two observing runs \citep{LVC_catalog_18}.  These observations have facilitated the placing of constraints on the departure of the true theory of gravity from Einstein's theory of general relativity \citep{LVC_testing_GR_19}, given insight into compact binary evolution \citep{LVC_BBH_populations_18}, and the physics of gamma ray bursts \citep{LVC_GWs_and_GRBs_17}.  They have also been used to place constraints on the equation of state of matter at high densities \citep{LVC_EoS_18}.

All of the gravitational wave detections thus far have been of compact binary coalescences.  An alternative gravitational wave source are \emph{continuous waves}---long lived nearly monochromatic signals, probably emitted by rotating or oscillating neutron stars.  


One way in which a steadily rotating neutron star could emit such a signal is if its mass distribution is not axisymmetric  with respect to the rotation axis.  Such a star would typically emit gravitational waves at twice its spin frequency, sourced by its non-zero mass quadrupole.  Such deformations are colloquially termed `mountains'.  See \citet{glampedakis_gualtieri_18} for a review.

A number of searches have been carried out for such signals.  No detections have been made, and upper limits have been placed on their strength; see \citet{LVC_allsky_isolated_O2_19}, \citet{LVC_known_pulsars_O1-O2_19} and \citet{LVC_SCo-X1-O2_19} for recent examples.  A prime target for such searches are the galactic low mass X-ray binaries (LMXBs).  As pointed out by Bildsten, while rotating rapidly, many in the interval $250$--$600$ Hz, the accretion to which they are subject could be expected to spin them up to significantly higher frequencies \citep{Bildsten1998}.  Bildsten suggested that this may be due to gravitational wave emission from a mountain, whose associated breaking torque scales as a steep (fifth) power of the spin frequency $\nu_{\rm s}$.  Bildsten calculated that relatively modest asymmetries in the moment of inertia tensor would be sufficient to halt the spin-up:
\be
\label{eq:bildsten_eqm_epsilon}
\epsilon \approx 5 \times 10^{-8} \left(\frac{\dot M}{10^{-9} \, M_\odot \, {\rm yr}^{-1}}\right)^{1/2}
\left(\frac{300 \, \rm Hz}{\nu_{\rm s}}\right)^{5/2} ,
\ee
where $\dot M$ is the accretion rate, and $\epsilon$ measures the fractional difference in the principal moments of inertia orthogonal to the spin axis, $\epsilon \equiv (I_{xx} - I_{yy}) / I_{zz}$ (assuming spin along $Oz$).

Bildsten proposed a specific mechanism through which such deformation might be built.  The mechanism exploited the fact that as mass elements in the solid crust are pushed downward through the arrival of newly accreted material, they undergo a series of nuclear reactions, giving the crust a layered structure.  These reactions have a weak temperature sensitivity, such that transitions to a given nuclear element occur a smaller distance below the crust's surface where the crust is hotter than the average.  This means that if the star's temperature is not axisymmetric with respect to the rotation axis, the mass distribution itself will inherit this non-axisymmetry, hence building the mountain.

This idea was explored in more detail by \citet{UCB}, hereafter UCB,  who showed that non-axisymmetries at the $\sim 10\%$ level in either the nuclear heating rate, or the composition, could provide the necessary asymmetry.   They showed that, regardless of its origin, a temperature asymmetry around the percent level could be enough to sustain the mountain. By fitting to their  results for a single capture layer (combining information from their figures 9 and 17), and allowing for the fact that there will be several capture layers, we have produced a simple numerical fit to their fractional temperature asymmetries $\delta T / T$,
\be
\label{eq:epsilon_T_UCB}
\epsilon \sim 5 \times 10^{-8} \frac{\delta T / T}{10^{-2}} ,
\ee
which we will use to convert our estimates of $\delta T / T$ below into estimates of mountain sizes.  However, as noted by UCB, a first-principles calculation of what sort of heating or compositional asymmetries might exist, and therefore what level of temperature asymmetry we might expect, is extremely challenging.  

In this paper we will explore one rather simple, alternative mechanism, for building a temperature asymmetry in a LMXB.  We will assume the star is threaded with a magnetic field, such that its thermal conductivity tensor is no longer isotropic.  This would naturally build a temperature asymmetry.  Such asymmetries have been considered in the galactic magnetar population, see e.g.\ \citet{Aguilera2008}.   The calculations presented here are thus similar to those performed for magnetars, with a few differences: i) We include heating from nuclear reactions; ii) We use a crustal equation of state appropriate for an accreted crust; iii) we exploit the likely weakness of the magnetic fields to carry out our calculation perturbatively.

This calculation is of interest for two reasons.  Firstly, it gives us a better idea of the likely gravitational wave detectability of the LMXBs.   Secondly, any detections can be used to place constraints on combinations of the various crustal parameters that appear in our calculations, including  the breaking strain and shear modulus, thermal conductivity, and the arrangement of the magnetic field, and the detailed composition of the crust.  The physics probed here is thus complementary to the aspects of the global equation of state probed by observations in binary neutron star coalescence.

Very recently, \citet{singh_etal_19} described a different mechanism for building temperature asymmetries in accreting magnetised neutron stars.  They made use of the physical shift in capture layers induced by the magnetic strains, and is quite different from the mechanism considered here.

The structure of this paper is as follows.  In Section \ref{sect:number} we give a rough estimate of the level of temperature asymmetry we can expect.  In Section \ref{sect:hydro} we describe the hydrostatic structure of our stellar model.  In Section \ref{sect:microphysics} we outline the various microphysical inputs we use.  In Section \ref{sect:thermal_background} we compute the thermal structure of our unmagnetised spherically symmetric stellar background.  In Section \ref{sect:perturbed_magnetic} we compute the perturbation in temperature induced when including an internal magnetic field.  In Section \ref{sect:perturbed_surface} we experiment with allowing for a non-spherical surface temperature distribution, as might be caused by non-spherical accretion.  In Section \ref{sect:improvements}, we describe some areas where our model might be improved.  Finally, in Section \ref{sect:summary} we summarise and discuss our results.

\section{A Rough estimate} \label{sect:number}

Before describing our model, we will first make a rough estimate of the importance of magnetic fields on the thermal conductivity, by making an estimate of the dimensionless \emph{magnetisation parameter}, as described in see \citet{yakovlev_urpin_80}.  This parameter  is defined as
\be
\omega \tau \equiv \frac{e B}{m_e^* c} \tau ,
\ee
where $e$ is the electronic charge and  $m_e^*$ the relativistic mass of the electrons, such that $\omega \equiv e B / (m_e^* c)$ is the gyrofrequency of the electrons, and $\tau$ their scattering timescale.  We will find that electron-phonon scattering is the dominant scattering mechanism, so that $\tau$ becomes a simple function of temperature; see equation  (\ref{eq:nu_ep}) below.  At the densities of interest, the electrons are highly relativistic, and it is straightforward to evaluate $m_*$ as a function of density and composition, leading to the result:
\be
\label{eq:omega_tau}
\omega \tau \approx 2 \times 10^{-4} \frac{B_9}{T_8} \left[\frac{A_{46}}{Z_{14} \rho_{12}} \frac{0.82}{1-X_n} \right]^{1/3} ,
\ee
where  we have scaled to the composition corresponding to a density $\rho_{12} = \rho / 10^{12}$ g cm$^{-3}$ for the accreted crust equation of state of \citet{HandZ90b}, i.e. atomic number $Z_{14} =Z/14$, mass number $A_{46} = A/46$ and free neutron fraction $X_n = 0.18$. Also, $T_8 = T / 10^8$ K, and $B_9 = B / 10^9$ G.    See Section \ref{sect:hydro} for more detail.

This sets the scale for the level of temperature asymmetry we might plausibly expect.  Given UCB found that percent-level temperature asymmetries are needed for gravitational wave emission to provide the LMXB braking mechanism, this indicates that internal magnetic fields approximately two orders of magnitude greater than the $B \sim 10^9$ G commonly believed to exist in LMXBs are required.  This argument may even under estimate the level of temperature asymmetry that might be produced.  As noted by \citet{Bildsten1998}, the predominant flux in an accreting LMXB crust will be radial, extending over the crustal thickness of $\sim 1$ km.  In contrast,  the thermal mountain requires a transverse (i.e.\ non-radial) temperature asymmetry over the longer lengthscale of the stellar circumference of $\sim 10$ km, introducing an aspect ratio of order $10$.

We take these considerations as motivation for carrying out a more detailed analysis of the actual temperature asymmetries that will be supported in our accreting star.

\section{Hydrostatic structure}
\label{sect:hydro}

We first need to fix the hydrostatic structure of our star.  We use the SLy equation of state in the core \citep{douchin_haensel_01}, and the equation of state of Haensel \& Zdunik, in  the crust \citep{HandZ90a, HandZ90b}, the latter being appropriate to a star with an accreted crust, assuming that the ashes of the nuclear burning of the accreted H and He are iron.  We employ Newtonian gravity, rather than general relativity, in computing the structure of the crust, as the inclusion of thermal transport and magnetic structure will then be easier in the later calculations.  However, to ensure we have a sensible overall stellar structure, we solve the full general relativistic Tolman-Oppenheimer-Volkov (TOV) equations in the stellar core, appending the Newtonian accreted crust on top, at a transition density of $1.25  \times 10^{13}$ g cm$^{-3}$, corresponding to the highest density of Table 2 of  \citet{HandZ90b}.

This gives us a star of overall mass $1.84 M_\odot$  and radius $11.6$ km.  The crust itself has mass $0.09 M_\odot$, and thickness $1.45$ km.   By comparing our hybrid TOV--Newtonian gravity calculation with the (correct) TOV-only calculation, we found that the error we make in stellar properties is rather modest, with our star having a slightly greater mass $\Delta M \approx 4 \times 10^{-3}  M_\odot$ and slightly greater radius $\Delta R \approx 400$ m than the TOV-only star.  

\begin{figure}
\begin{center}
\begin{minipage}[c]{\linewidth}
\includegraphics[width=\textwidth]{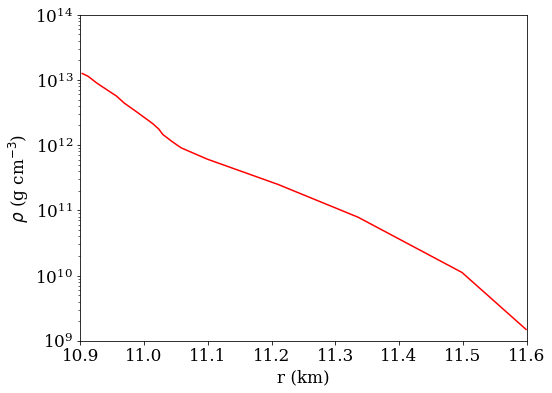}
\caption{ Density versus radius for our accreted crust.   \label{fig:rho_v_r} }
\end{minipage}
\end{center}
\end{figure}

In the thermal calculations that follow, we will restrict our computational domain to the range of densities where Haensel \& Zdunik have computed the heat release, so that our inner boundary is defined by the density and radius:
\be
\rho_{\rm IB} = 1.25 \times 10^{13} \, {\rm g \, cm}^{-3} , \hspace{5mm} R_{\rm IB} = 10.903 \, {\rm km} ,
\ee
and at the outer boundary:
\be
\rho_{\rm OB} = 1.49 \times 10^{9} \, {\rm g \, cm}^{-3} , \hspace{5mm} R_{\rm OB} = 11.60 \, {\rm km} .
\ee
A plot of the variation of density versus radius over our computational domain is given in Figure  \ref{fig:rho_v_r}.

\section{Microphysical inputs}
\label{sect:microphysics}

Before computing the thermal structure of our star, we first need to fix various pieces of microphysics.  For the most part, we follow the choices of UCB, as described briefly below.

\subsection{Accretion heating  } 
\label{sect:accretion}

As noted in Section \ref{sect:hydro}, we use the crustal equation of state of Haensel \& Zdunik, such that our crust is composed of layers of nuclear matter of a given nuclear species with mass and atomic numbers $(A, Z)$, with free neutrons making up a fraction $X_{\rm n}$ of the baryons, as given in Tables 1 and 2 of \citet{HandZ90a}.  Haensel \& Zdunik also give the heat energy deposited, per accreted baryon, at each of the transition layers between these shells of given $(A, Z)$.  The total heat deposited is $1.33$ MeV per accreted baryon.

Strictly, the transition layer will have a finite thickness that depends upon a balance between the accretion rate and the relevant nuclear reaction rate, with the heat being released over the transition layer, as computed in detail in UCB.  We take a simpler approach, and treat the transitions as infinitely sharp, and instead smear the heat released at each transition over the shell of constant $(A, Z)$.  This is less accurate than the analysis of UCB, where the precise variation of composition over the transition layer was needed to compute the shifts in position of the layers when a temperature perturbation was introduced.  This approximation should be fine for us, as we do not need to resolve such fine detail, and is in fact more accurate that heating prescriptions used elsewhere, e.g. \citet{brown_00, Brown2009}.

\subsection{Neutrino cooling  }
\label{sect:neutrinos}

We follow UCB in modelling the neutrino cooling in the crust as due to bremstrahlung interactions, as computed by \citet{HKY1996}, giving a cooling rate per unit time per unit volume of
\be
\label{eq:Q_dot_nu}
\dot Q_\nu  = 3.229 \times 10^{11} \rho_{12} T_8^6 \frac{Z^2}{A} (1-X_n) \, {\rm erg \, s}^{-1} \, {\rm cm}^{-3} .
\ee

\subsection{Thermal conductivity }
\label{sect:conductivity}

Using the results of \citet{yakovlev_urpin_80}, we model the thermal conductivity $K$ as due purely to the mobility of the electrons, such that 
\be
K = \frac{\pi^2 k_{\rm B}^2 n_e}{3 m_e^*} T \tau ,
\ee
where $k_{\rm B}$ is Boltzmann's constant and $n_e$ the electron density.  We allow for electron-phonon and electron-impurity scattering, such that
\be
\nu = \frac{1}{\tau} = \nu_{ep} + \nu_{eQ} ,
\ee
where $\nu_{ep}$ and $\nu_{eQ}$ are the scattering rates for these two processes.  For electron-phonon scattering we have
\be
\label{eq:nu_ep}
\nu_{ep} = \frac{13 e^2 k_{\rm B} T}{\hbar^2 c}   ,
\ee
which is a function only of temperature; see e.g.  \citet{yakovlev_urpin_80} and the appendix of \citet{Brown2009}.  For electron-impurity scattering we have
\be
\nu_{eQ} = \frac{4 \pi Q_{\rm imp} e^4 n_{\rm ion}}{p_{\rm F}^2 v_{\rm F} } \Lambda_{\rm imp}  ;
\ee
again, see  \citet{yakovlev_urpin_80} and  \citet{Brown2009}.  In this equation,  $p_{\rm F}$ is the electron Fermi momentum, $v_{\rm F}$ the Fermi velocity, $n_{\rm ion}$ the ion density, $\Lambda_{\rm imp} \sim 1$ the Coulomb parameter.  Also, $Q_{\rm imp}$ is the impurity parameter, a measure of the deviation of the atomic number from its average value:
\be
Q_{\rm imp} \equiv \frac{1}{n_{\rm ion}} \sum_i n_i ( Z_i - \langle Z \rangle )^2 ,
\ee
where the sum $i$ is over all the different species of ions, with atomic number $Z_i$ and mean atomic number $\langle Z \rangle$.

The crustal equation of state of Haensel \& Zdunik that we employ assumes only one type of nucleus $(A, Z)$ is present at any given density, so strictly $Q_{\rm imp}=0$ for such a composition.  We will nevertheless allow for $Q_{\rm imp} \neq 0$, as it is likely that real crusts will not be perfectly pure, although recent modelling of transiently accreting LMXBs point to rather low values of $Q_{\rm imp}$, of order unity \citep{Brown2009}.

In the presence of a magnetic field, the thermal conductivity becomes anisotropic, and must be represented as a tensor \citep{yakovlev_urpin_80}.  This will be described in Section \ref{sect:perturbation_equations}  below.

\section{Thermal structure of the unmagnetised spherical background}
\label{sect:thermal_background}

We next compute the thermal structure of our steadily-accreting star, in the absence of a magnetic field, so that the temperature profile is spherically symmetric and unchanging in time.  By energy conservation, the heat flux $\bf F$ is related to the net rate of heat energy generation per unit volume $\dot Q$ as:
\be
\nabla \cdot {\bf F} = \dot Q .
\ee
The heat flux is related to the temperature $T$ via Fourier's law:
\be
{\bf F} = - K \nabla T .
\ee
Writing these out for our spherically symmetric system we obtain two coupled ordinary differential equations (ODEs) for $F = |{\bf F}|$ and $T$, with independent variable radius $r$:
\begin{eqnarray}
\label{eq:dFbdr}
\frac{dF}{dr} &=& \dot Q - \frac{2}{r} F ,\\
\label{eq:dTbdr}
\frac{dT}{dr} &=& - \frac{1}{K} F .
\end{eqnarray}

We use the same boundary conditions as employed in UCB.  The temperature at the outer boundary is fixed by the ratio of the local accretion rate $\dot m$ to the local Eddington accretion rate $\dot m_{\rm Edd}$,
\be 
\label{eqT_OB}
T_{\rm OB} = 5.3 \times 10^8 {\, \rm K \,}  \left(\frac{\dot m}{\dot m_{\rm Edd}}\right)^{2/7} ,
\ee
as computed by \citet{Schatz_1999}.  We assume that the core is isothermal with a temperature equal to that of the inner boundary. The inner boundary condition then comes from equating the flux flowing into the core to the neutrino luminosity $L_{\rm core}$ of the core:
\be
-4 \pi R_{\rm IB}^2 F_{\rm IB} = L (T_{\rm IB}) ,
\ee
where we use the core neutrino modified Urca luminosity of \citet{SandT}:
\be
L = 5.31 \times 10^{31} {\, \rm  erg \, s}^{-1} \, \left(\frac{M}{M_\odot}\right) \left(\frac{\rho_{\rm nuc}}{\rho}\right)^{1/3}  
T_8^8 \exp\left(-\frac{\Delta}{k_{\rm B} T_{\rm IB}} \right) ,
\ee
where $\rho_{\rm nuc}$ is nuclear density.  We consider two cases.  The first is that of a normal core, for which $\Delta = 0$.  The second is a superfluid core, for which $\Delta \sim 1$ MeV.  For the sorts of temperatures we will encounter, this will make the exponential factor very small ($\exp(-\Delta/ k_{\rm B} T_{\rm IB}) \sim 10^{-6}$), and we  will simply set
\be
F_{\rm IB} = 0 
\ee
for the superfluid case.  This corresponds to the `strong superfluidity' approximation used by \citet{brown_00}.

We then proceeded to numerically solve the coupled ODEs (\ref{eq:dFbdr}) and (\ref{eq:dTbdr}).  These form a boundary value problem.  We solved the equations using the python ODE solver \texttt{ODEint}, dividing the computational domain into $N$ sub-intervals, and integrating inwards from the outer boundary to the inner boundary, using the known value for $T_{\rm OB}$,  and making a guess for the unknown value of the flux $F_{\rm OB}$.  The python root solving package \texttt{brentq} was used to find the correct $F_{\rm OB}$ such that the inner boundary condition was also satisfied, thereby obtaining $T(r)$ and $F(r)$.  

We performed various consistency tests of our results.  These included energy conservation, evaluation of the left hand side of the ODEs by stencil-based finite differencing of $F(r)$ and $T(r)$, and convergence  of the results with respect to the number of sub-intervals $N$.  In all cases, sensible results were obtained, e.g. energy conservation was typically satisfied to better than a fractional accuracy of $10^{-5}$.

\begin{figure*}
  \begin{subfigure}[b]{0.45\textwidth}
    \includegraphics[width=\textwidth]{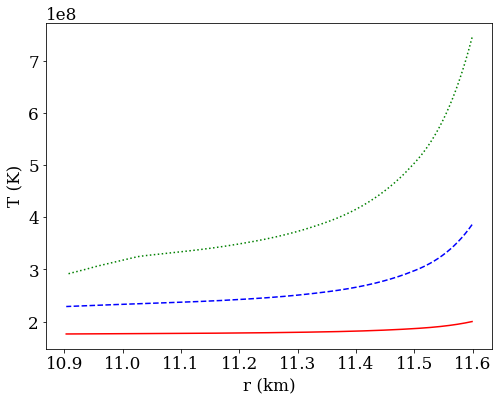}
  \end{subfigure}
  \begin{subfigure}[b]{0.45\textwidth}
    \includegraphics[width=\textwidth]{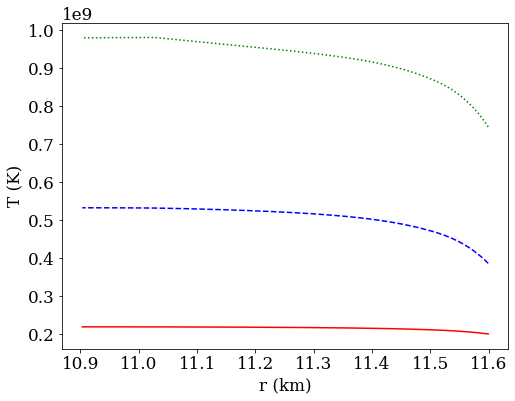}
  \end{subfigure}
\caption{Left: Temperature versus radius in the crust of a star with a normal fluid core,  with impurity parameters $Q_{\rm imp}=1$, for accretion rates of  $10^{-8} M_\odot$ yr$^{-1}$ (green, dotted), $10^{-9} M_\odot$ yr$^{-1}$ (blue, dashed),  and $10^{-10} M_\odot$ yr$^{-1}$ (red, solid).   Right: Same, but for a star with a superfluid core.   \label{fig_T_v_r}}  
\end{figure*}

In Figure  \ref{fig_T_v_r}, we give plots of temperature versus radius, for impurity parameter $Q_{\rm imp}=1$, and for three different accretion rates, for a normal core (left hand plot) , and a superfluid core (right hand plot).  We find that for $Q_{\rm imp}=1$ and the temperatures obtained, the conductivity is dominated by electron-phonon scattering, with $\nu_{ep} / \nu_{eQ} \gtrsim 100$, i.e. we are firmly in the electron-phonon scattering regime.  In the case of a normal core, the temperature gradient is postive, corresponding to an inwards flow of heat throughout the entire crust.  In the case of a superfluid core, the temperature is slightly higher, and over much of the crust the temperature gradient is negative, corresponding to a positive heat flux.  This can be attributed to the superfluid inner boundary condition of zero heat flux, so heat can no longer be lost through conduction into the core.

\section{Perturbation in Thermal structure caused by magnetic field}
\label{sect:perturbed_magnetic}

\subsection{Derivation of the perturbation equations}
\label{sect:perturbation_equations}

We now perturb the thermal solutions found in Section \ref{sect:thermal_background} through the addition of a magnetic field $\bf B$, to give a new steady-state (constant in time) solution.   This amounts to linearising in the magnetisation parameter $\omega \tau$.  For temperatures $T \sim 10^8$ K typical of LMXBs, equation (\ref{eq:omega_tau}) indicates this should be safe for internal magnetic fields strengths $B \lesssim 10^{12}$ G, i.e.\ valid for field strengths up to three orders of magnitude larger than the external field strengths  $\sim 10^9$ G normally assumed for LMXBs.

Perturbing the energy conservation equation gives
\be
\nabla \cdot {\bf \delta F} = \delta \dot Q .
\ee
We will assume that this addition of the magnetic field produces a change in the net heating rate only via the dependence of $\dot Q$ on temperature (via the temperature dependence of the neutrino cooling; see equation (\ref{eq:Q_dot_nu})), so this can be written as:
\be
\label{eq:pert_energy_cons}
\nabla \cdot {\bf \delta F} = \frac{d\dot Q}{dT} \delta T  . 
\ee
Given the spherical symmetry of our background, we will use spherical harmonics to describe the dependance of all perturbed quantities on the spherical polars angles $(\theta, \phi)$, e.g.
\be
\label{eq:delta_T_decomp}
\delta T(r, \theta, \phi) = \sum_{lm} \delta T_{lm}(r) Y_{lm}(\theta, \phi) .
\ee
Clearly, only a polar (as opposed to axial) perturbation in the heat flux will couple to $\delta T$, so we decompose $\bf \delta F$ in purely polars terms:
\be
\label{eq:delta_F_decomp}
{\bf \delta F} = \sum_{lm} U_{lm}(r) {\bf \hat r} Y_{lm} + V_{lm}(r) \nabla Y_{lm} ,
\ee
where $\bf \hat r$ is the radial unit vector.  Inserting equations (\ref{eq:delta_T_decomp}) and (\ref{eq:delta_F_decomp}) into equation (\ref{eq:pert_energy_cons}) we obtain
the first ODE describing the perturbed thermal structure:
\be
\label{eq:dUlmbdr}
\frac{dU_{lm}}{dr} = -\frac{2}{r} U_{lm} + \frac{l(l+1)}{r^2} V_{lm} + \frac{d\dot Q}{dT} \delta T_{lm} .
\ee
Our remaining equations come from the  equation relating the flux and temperature gradient in the presence of a magnetic field, as computed by \citet{yakovlev_urpin_80}.  For a magnetic field ${\bf B} = B {\bf b}$, they show:
\be
{\bf F} = - K {\bf b} ({\bf b} \cdot \nabla T) - \frac{K}{1+ (\omega\tau)^2} {\bf b} \times (\nabla T \times {\bf b}) 
- \frac{K \omega\tau}{1 + (\omega\tau)^2} {\bf b} \times \nabla T .
\ee
For $\omega \tau = 0$ this reduces to the scalar conductivity of Section \ref{sect:thermal_background}.  It follows that for $\omega\tau \ll 1$ we can linearise in $\omega \tau$ and $T$, which leads to:
\be
\label{eq:delta_F}
{\bf \delta F} = -\frac{dK}{dT} \delta T \nabla T - K \nabla \delta T -  K \varpi\tau {\bf B} \times \nabla  T ,
\ee
where for convenience we have introduced the symbol
\be
\varpi \equiv \frac{e}{m_e^* c} ,
\ee
the `gyrofrequency per unit magnetic field strength'.  

We see that in this weak magnetic field limit, the effect of the field is to deflect a piece of the thermal flux `sideways', i.e. orthogonally with respect to the local magnetic field and temperature gradients.  Furthermore, given the spherical symmetry of the background thermal profile, the vector $\nabla T$ is poloidal; it follows that if one decomposes the magnetic field into poloidal and toroidal components, such that ${\bf B} = {\bf B_{\rm pol}} + {\bf B_{\rm tor}}$,     only the toroidal component affects the heat flux, i.e. ${\bf B_{\rm pol}} \times \nabla T = 0$.  

Given that we need only consider toroidal fields, we can write $\bf B$ in terms of a scalar function $\psi$:
\be
{\bf B} = \nabla \times  (\psi {\bf r}) .
\ee
Inserting this into (\ref{eq:delta_F})  and decomposing in terms of spherical harmonics leads to the ODE 
\be
\label{eq:dTlmbdr}
\frac{d\delta T_{lm}}{dr} = -\frac{1}{K} \frac{dK}{dT} \frac{dT}{dr} \delta T_{lm} - \frac{1}{K} U_{lm} ,
\ee
together with the algebraic equation
\be
\label{eq:Vlm}
V_{lm} = -K \delta T_{lm} + K \varpi\tau r \frac{dT}{dr} \psi_{lm} .
\ee
Once a magnetic field $\bf B$ is specified, the two ODES of equations (\ref{eq:dUlmbdr}) and (\ref{eq:dTlmbdr}), together with equation (\ref{eq:Vlm}), form a set of three equations for the unknowns $\delta T_{lm}, U_{lm}, V_{lm}$.  We use the same boundary conditions as in UCB, with zero temperature perturbation at both the inner and outer boundaries
\be
\label{eq:perturbed_BCs}
\delta T(R_{\rm IB}) = \delta T(R_{\rm OB}) = 0 ,
\ee
on account of the significantly higher conductivities in the core and ocean.

We will specialise to the $l=m=2$ component, as this is the one that would give rise to the time-varying mass quadrupole commonly targeted in gravitational wave searches, if our star were to rotate about the $z$-axis.  We do not include rotation in our analysis, and so we neglect the small effect of rotation on the mass distribution of our star.

In Section \ref{sect:perturbed_surface} we will modify the outer boundary condition, to allow for a temperature asymmetry at the stellar surface, as might be produced by non-isotropic accretion.

\subsection{Choice of magnetic field}
\label{sect:magnetic_field}

We need to specify a form for our toroidal magnetic field.  We are only interested in the $l=m=2$ spherical harmonic.  We will follow the choice made in \citet{Pons07} and take
\be
\psi_{22}(r) = C [(r-R_{\rm IB})(r-R_{\rm OB})]^2 ,
\ee
with $C$ constant, such that the magnetic field is only non-zero within our computational domain.  Such a choice gives a magnetic field
\begin{multline}
{\bf B}     = C \frac{1}{4} \sqrt{\frac{15}{2\pi}} [(r-R_{\rm IB})r-R_{\rm OB})]^2 \\  
   \times  (-\sin\theta \sin2\phi {\bf e_\theta} + \sin\theta \cos\theta \cos2\phi {\bf e_\phi} ) .
\end{multline}

Clearly, the magnetic field strength $B$ is a function of position.  Will will report the strength at the (fairly arbitrarily chosen) point $\theta = \pi/2$, $\phi = \pi/4$, and half way between $r=R_{\rm IB}$ and $r= R_{\rm OB}$, when describing the strength of a  given magnetic configuration.

\subsection{Results: without shallow heating}
\label{sect:results_without_SH}

We numerically solved the coupled ODEs of equations (\ref{eq:dUlmbdr}), (\ref{eq:dTlmbdr}) and (\ref{eq:Vlm}), using the outer boundary conditions of equation (\ref{eq:perturbed_BCs}), and the magnetic field  described in Section \ref{sect:magnetic_field}.  We used the same combination of Python functions to solve via the shooting method as was described in Section \ref{sect:thermal_background}.  As a check on our numerics, we again used stencil-based finite differencing to check that the two ODEs for $U_{22}$ and $\delta T_{22}$ were satisfied, and that our solutions were converted with respect to the number of subintervals into which we divided our computational domain.

\begin{figure*}
  \begin{subfigure}[b]{0.45\textwidth}
    \includegraphics[width=\textwidth]{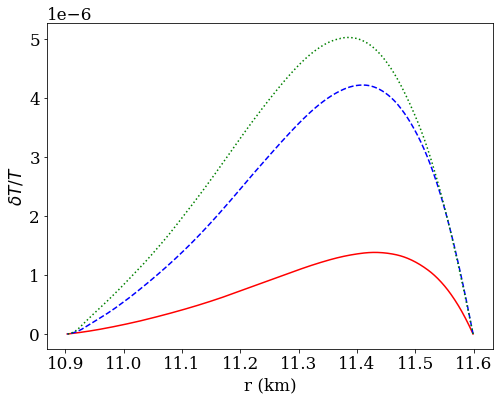}
     \end{subfigure}
  \begin{subfigure}[b]{0.45\textwidth}
    \includegraphics[width=\textwidth]{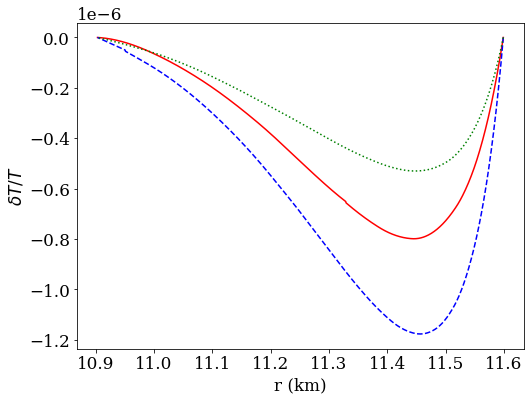}
  \end{subfigure}
\caption{Left: Fractional temperature perturbation $\delta T(r)/T(r)$ versus radius, for  a star with a normal fluid core, a magnetic field strength $B= 10^9$ G, an  impurity parameter $Q_{\rm imp}=1$ ,  for the accretion rates of  $10^{-8} M_\odot$ yr$^{-1}$ (green, dotted), $10^{-9} M_\odot$ yr$^{-1}$ (blue, dashed),  and $10^{-10} M_\odot$ yr$^{-1}$ (red, solid).  Right: same, but for a star with a superfluid core.  \label{fig_dT_over_T_v_r}}  
\end{figure*}

\begin{figure*}
  \begin{subfigure}[b]{0.45\textwidth}
    \includegraphics[width=\textwidth]{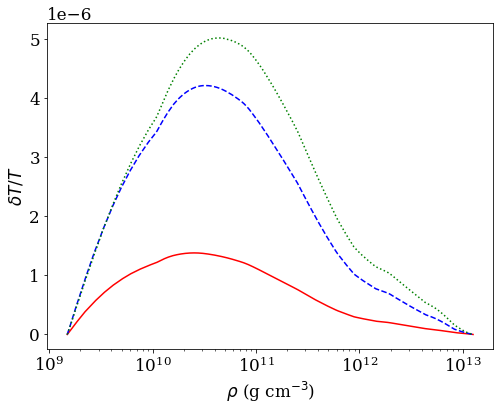}
     \end{subfigure}
  \begin{subfigure}[b]{0.45\textwidth}
    \includegraphics[width=\textwidth]{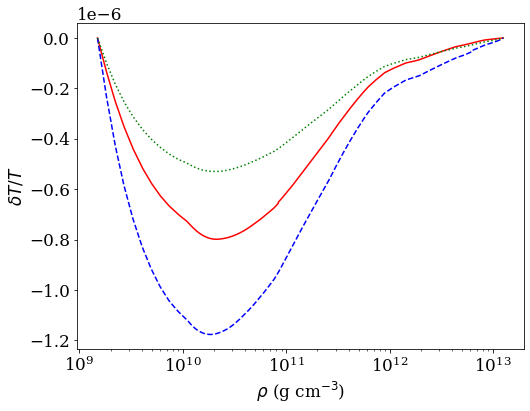}
  \end{subfigure}
\caption{  Same as Figure \ref{fig_dT_over_T_v_r}, but now we plot the fractional temperature as a function of density, not radius.   \label{fig_dT_over_T_v_rho}}  
\end{figure*}

In Figure \ref{fig_dT_over_T_v_r} we plot the fractional temperature perturbation $\delta T/T$ as a function of radius (dropping the subscript $(22)$ from this point on), for a normal core (left hand plot) and a superfluid core (right hand plot).  We have set $Q_{\rm imp}=1$ and $B = 10^9$ G.  The corresponding plots for $\delta T/T$ as a function of density are given in Figures \ref{fig_dT_over_T_v_rho}.  We see typical values of $\delta T/T$ of a few times  $10^{-6}$ for both sorts of core, peaking towards the outer crust, at densities of a few times $10^{10}$ g cm$^{-3}$.    

There are however a few significant differences between the normal and superfluid core results.  The overall sign of the perturbation is different in the two cases, and the ordering in terms of accretion rate is different, with the largest (in magnitude) fractional temperature perturbation occurring for the highest accretion rate in the normal case, and for the smallest accretion rate in the superfluid case.

\begin{figure*}
  \begin{subfigure}[b]{0.45\textwidth}
    \includegraphics[width=\textwidth]{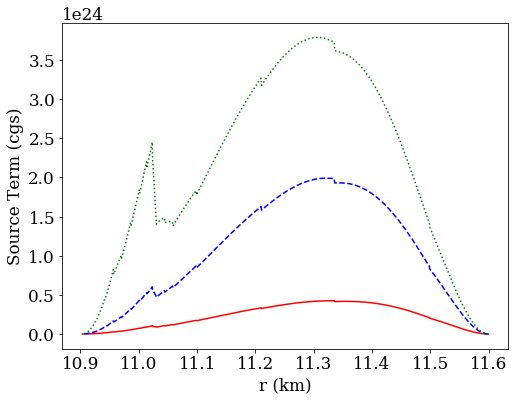}
     \end{subfigure}
  \begin{subfigure}[b]{0.45\textwidth}
    \includegraphics[width=\textwidth]{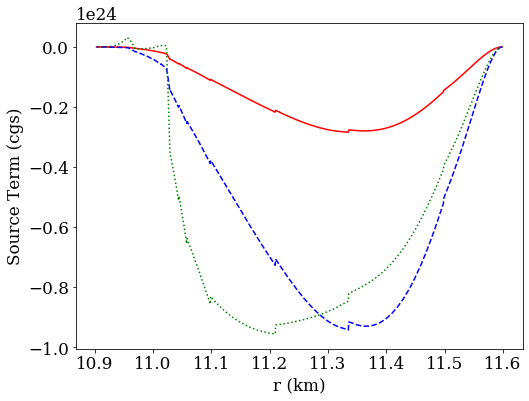}
  \end{subfigure}
\caption{ Same as Figure \ref{fig_dT_over_T_v_r}, but now we plot the source term, as defined in equaiton (\ref{eq:S}).     \label{fig_source_v_r}}  
\end{figure*}

One can attempt to gain a little insight into this behaviour by considering the source term in our problem, i.e. the final term in equation (\ref{eq:Vlm}):
\be
\label{eq:S}
S \equiv K \varpi\tau r \frac{dT}{dr} \psi_{lm} .
\ee
We plot this term in Figure \ref{fig_source_v_r}.  We can see that the sign of this term is proportional to the gradient of the background temperature.  This gradient was of opposite sign for the normal and superfluid cases (see Figure \ref{fig_T_v_r}), immediately accounting for the overall sign difference in the perturbations.  

To try and understand the dependence of our results on accretion rate, we need to isolate those factors in $S$ that depend upon the background temperature.  We have
\be
S \sim K \tau \frac{dT}{dr} \sim T \tau^2 \frac{dT}{dr} \sim \frac{1}{T} \frac{dT}{dr} ,
\ee
where we have used the scalings $K \sim T \tau$ and, specific to the regime where electron-phonon scattering dominates the conductivity, $\tau \sim 1/T$.  We therefore see that the source term is independent of $T$, in the sense it is invariant under simple re-scalings.  For this reason, there is no \emph{apriori} reason to expect a simple scaling of our perturbation results with the overall accretion rate, with the exact form of the gradient of $T$ playing an important role.  This sensitivity to the exact form of the background results presumably accounts for the rather different variations of the temperature perturbations with accretion rate, when comparing the normal and superfluid cases.

\begin{figure*}
  \begin{subfigure}[b]{0.45\textwidth}
    \includegraphics[width=\textwidth]{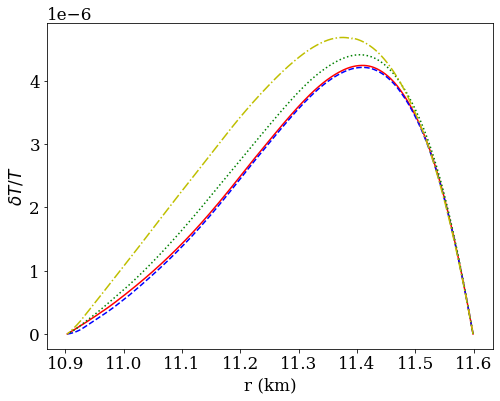}
     \end{subfigure}
  \begin{subfigure}[b]{0.45\textwidth}
    \includegraphics[width=\textwidth]{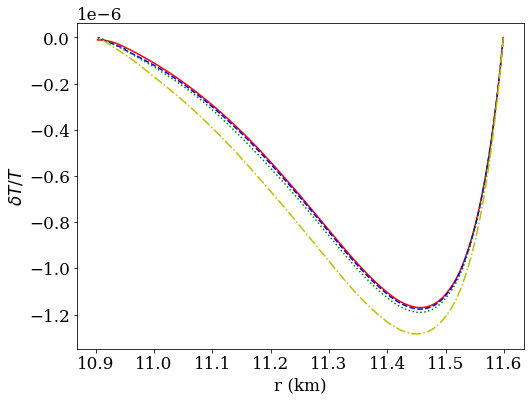}
  \end{subfigure}
\caption{ Left: Fractional temperature perturbation $\delta T(r)/T(r)$ versus radius, for  a star with a normal fluid core, a magnetic field strength $B= 10^9$ G, an accretion rate $\dot M = 10^{-9} M_\odot$ yr$^{-1}$, for a range of impurity parameters $Q_{\rm imp} = 0.1$ (red, solid), $1.0$ (blue, dashed), $10$ (green, dotted) and $100$ (yellow, dot-dash).   Right: same, but for a star with a superfluid core.  \label{fig_dT_over_T_v_r_Q}}  
\end{figure*}

In Figure  \ref{fig_dT_over_T_v_r_Q} we show how the form of the fractional temperature perturbation $\delta T / T$  depends upon the value of the impurity parameter $Q_{\rm imp}$, for a fixed accretion rate of $10^{-9} M_\odot$ yr$^{-1}$, for both normal and superfluid cores.  In both cases, for $Q_{\rm imp} \lesssim 10$, there is almost no dependence of the results on the value of $Q_{\rm imp}$, consistent with the conductivity being dominated by electron-phonon scattering.  Only for $Q_{\rm imp} \gtrsim 10$ do the results differ significantly from the $Q_{\rm imp} \ll 1$ results, with larger $Q_{\rm imp}$ values giving rise to very slightly larger perturbations in the temperature.  Even here the dependence is rather weak.  We therefore conclude that for reasonable ranges in $Q_{\rm imp}$, our results depend only weakly upon the impurity parameter.

It would be useful to produce a simple formula to summarise our numerical results, writing the  size of the temperature perturbation $\delta T / T$ as a linear function of magnetic field strength.  This could then be combined with the fitting formula of equation (\ref{eq:epsilon_T_UCB})  above, which writes the ellipticity $\epsilon$ as a linear function of $\delta T / T$, which was obtained by fitting to the results of UCB, to give a formula relating $\epsilon$ and $B$.  Note, however, that while equation (\ref{eq:epsilon_T_UCB}) is roughly independent of accretion rate, our results do have some $\dot M$-dependence, which does not appear to be of a form that can be conveniently expressed as a simple scaling.  We will therefore pick out our $\dot M = 10^{-9} M_\odot$ yr$^{-1}$ results in writing down our formula, with the understanding that the results can change by a factor of a few, depending upon the accretion rate, as per Figure \ref{fig_dT_over_T_v_rho}.  Also, our temperature perturbation is a function of position within the crust.   According to the analysis of UCB, it is the deeper capture layers that contribute most to the formation of the mass quadrupole.  We will therefore pick out the temperature perturbation at the relatively deep $\rho = 10^{12}$ g cm$^{-3}$ location as a representative value.   With these choices, for stars with normal cores we obtain
\be
\label{eq:deltaT_over_T_normal_B}
\frac{\delta T}{T} \sim 1 \times 10^{-6} B_9  \Rightarrow \epsilon \approx 5 \times 10^{-12}  B_9 ,
\ee 
where we have used equation (\ref{eq:epsilon_T_UCB}) to obtain the second equation.  Similarly, for stars with superfluid cores we obtain
\be
\label{eq:deltaT_over_T_SF_B}
\frac{\delta T}{T} \sim 2 \times 10^{-7} B_9  \Rightarrow \epsilon \approx 1 \times 10^{-12} B_9 .
\ee 

We will discuss these results in Section \ref{sect:summary}, but for now we will note the following.  The above formulae indicate that very large magnetic fields are necessary to produce ellipticities at the level of that of equation (\ref{eq:bildsten_eqm_epsilon}).  Taking the scalings literally, fields at the level of $\sim 10^{13}$ G are indicated.  Note, however, that such large fields push us out of the $\omega \tau \ll 1$  regime where the effect of the magnetic field on the thermal conductivity can safely be treated using out perturbative scheme (see equation (\ref{eq:omega_tau})), and so should be viewed with caution.

\subsection{Results: with shallow heating}
\label{sect:results_with_SH}

\begin{figure*}
  \begin{subfigure}[b]{0.45\textwidth}
    \includegraphics[width=\textwidth]{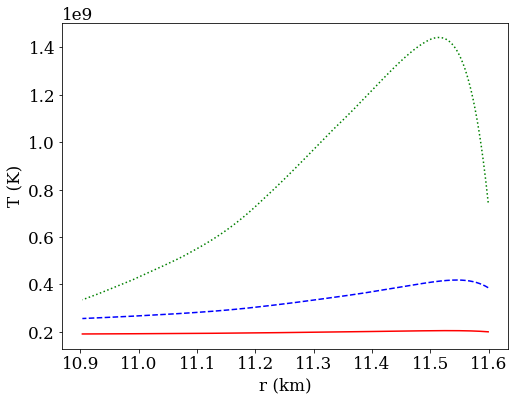}
     \end{subfigure}
  \begin{subfigure}[b]{0.45\textwidth}
    \includegraphics[width=\textwidth]{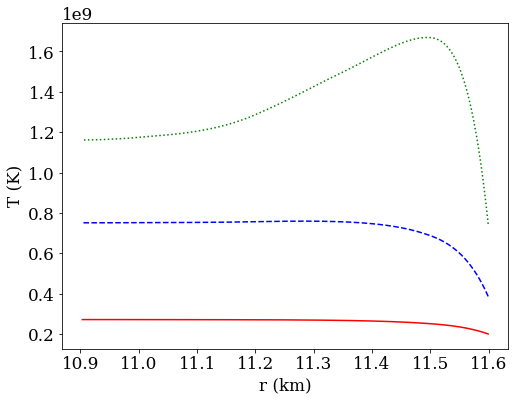}
  \end{subfigure}
\caption{ Same as Figure \ref{fig_T_v_r}, but now with shallow crustal heating.  \label{fig_T_v_r_SCH}}  
\end{figure*}

\begin{figure*}
  \begin{subfigure}[b]{0.45\textwidth}
    \includegraphics[width=\textwidth]{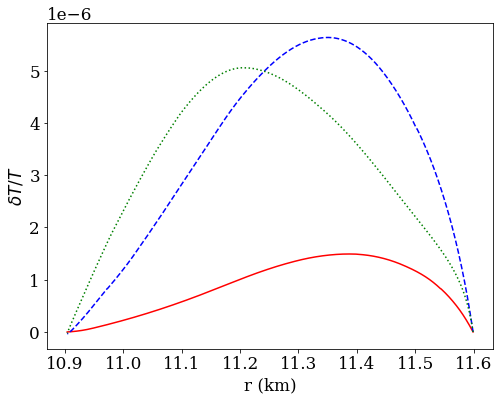}
     \end{subfigure}
  \begin{subfigure}[b]{0.45\textwidth}
    \includegraphics[width=\textwidth]{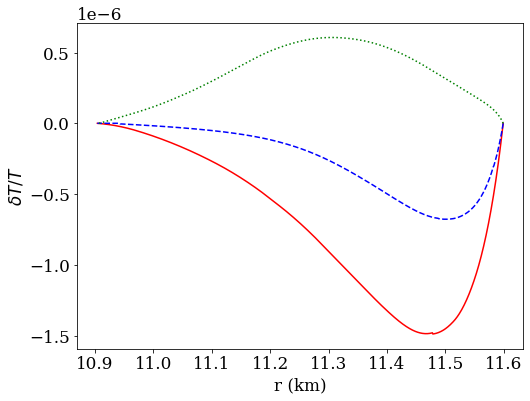}
  \end{subfigure}
\caption{  Same as Figure \ref{fig_dT_over_T_v_r}, but now with shallow crustal heating.  \label{fig_dT_over_T_v_r_SCH}}  
\end{figure*}

From the analysis of the time-dependent light curves of transiently accreting LMXBs, there is evidence for an additional source of heating, over and above the $\sim 1.33$ MeV per nucleon associated with known reactions; see e.g.\ \citet{Deibel2015}.  With this in mind, we repeated the above analysis, adding in an extra $2$ MeV per nucleon in each of the three lowest density compositional shells of \citet{HandZ90a, HandZ90b}, defined by the densities $\rho = 1.49 \times 10^9$ g cm$^{-3}$,  $\rho = 1.11 \times 10^{10}$ g cm$^{-3}$ and $\rho = 7.85 \times 10^{10}$ g cm$^{-3}$.  We are therefore adding an extra $6$ MeV per nucleon of heat energy.   The background temperatures are shown in Figure \ref{fig_T_v_r_SCH}, and the temperature perturbations in Figure \ref{fig_dT_over_T_v_r_SCH}. 

Again using the temperature perturbation at $\rho = 10^{12}$ g cm$^{-3}$ for the $\dot M = 10^{-9} M_\odot$ yr$^{-1}$ case as indicative, for stars with normal cores we have
\be
\label{eq:deltaT_over_T_normal_B_SCH}
\frac{\delta T}{T} \sim 2 \times 10^{-6}  B_9  \Rightarrow \epsilon \approx 1 \times 10^{-11} B_9 ,
\ee 
while for stars with superfluid cores we obtain
\be
\label{eq:deltaT_over_T_SF_B_SCH}
\frac{\delta T}{T} \sim 4 \times 10^{-8} B_9  \Rightarrow \epsilon \approx 2 \times 10^{-13} B_9 .
\ee

\section{Non-spherical surface temperatures}
\label{sect:perturbed_surface}

In Sections \ref{sect:results_without_SH} and \ref{sect:results_with_SH} we have calculated the level of temperature asymmetry that can be expected in a magnetised LMXB, both without and with shallow crustal heating.  Clearly, extremely large internal toroidal magnetic fields would be required to produce the percent-level temperature asymmetries that UCB find are needed to generate mass quadrupoles large enough to halt spin-up.  In this section we will therefore explore an alternative possible source  of asymmetry, namely a non-zero and non-spherical surface temperature perturbation, such that we replace $\delta T_{\rm OB} = 0$ with $\delta T_{\rm OB} = \delta T_{\rm OB}(\theta, \phi)$.

Such a non-spherical surface temperature could arise if the local accretion rate $\dot m$ onto the surface is non-spherical, such as might be caused by the funnelling of the accretion flow by an \emph{external} magnetic field.  We can encode this asymmetry into a dimensionless function $\alpha$ such that
\be
\dot m = \alpha(\theta, \phi) \langle \dot m \rangle ,
\ee
where $\langle \dot m \rangle$ is the surface-average local accretion rate.  Specialising to the relevant quadrupolar piece, we can write:
\be
\delta \dot m = \alpha_{22} Y_{22}  \langle \dot m \rangle ,
\ee
where $\alpha_{22}$ is a constant (i.e. independent of $(\theta, \phi)$).   Note that an external \emph{dipolar} ($l=m=1$) magnetic field would naturally produce such an asymmetry, with the flow being directed towards the two diametrically opposed polar caps.

Our surface temperature boundary condition for the non-magnetised background solution was of the form $T_{\rm OB} \propto \dot m^{2/7}$, so we have a corresponding temperature perturbation
\be
\delta T_{\rm OB} = \frac{2}{7} T_{\rm OB}  \alpha_{22} Y_{22} .
\ee
If we write $\alpha_{22}$ in amplitude-phase form, such that
\be
\alpha_{22} = |\alpha_{22}| e^{-2 i \Delta \phi} ,
\ee
we can then have
\be
\delta T_{\rm OB} = \frac{2}{7} |\alpha_{22}|   T_{\rm OB}  Y_{22}(\theta, \phi-\Delta \phi) .
\ee
This shows that $\Delta \phi$ represents the relative location of the surface hot spots relative to the $\phi = 0$ axis that defines the symmetry axis of the internal toroidal magnetic field.  We will look only at real $\alpha_{22}$ in our analysis.  Now $\alpha_{22}$ is playing the role of our perturbation parameter, with $\alpha > 0$ corresponding to $\Delta\phi = 0$, and $\alpha < 0$ corresponding to $\Delta\phi = \pi/2$.  Results for intermediate angles could be obtained by suitably combining linear combinations of $m=+2$ and $m=-2$ solutions.

\begin{figure*}
  \begin{subfigure}[b]{0.45\textwidth}
    \includegraphics[width=\textwidth]{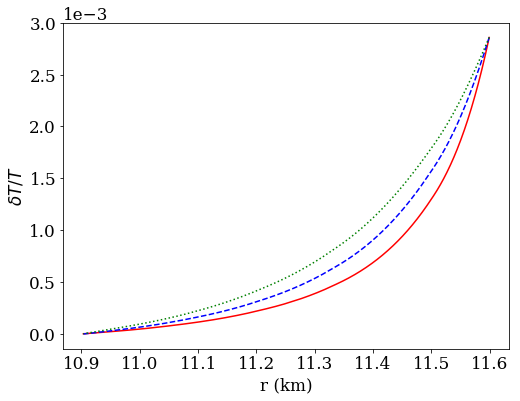}
     \end{subfigure}
  \begin{subfigure}[b]{0.45\textwidth}
    \includegraphics[width=\textwidth]{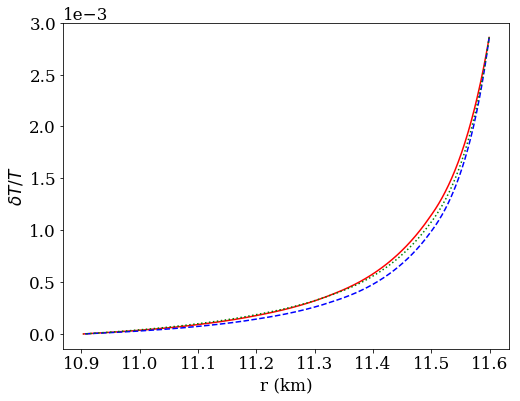}
  \end{subfigure}
\caption{  Left: Fractional temperature perturbation $\delta T(r)/T(r)$ versus radius, for  a star with a normal fluid core, zero internal  magnetic field strength, surface accretion asymmetry $\alpha = 10^{-2}$, an  impurity parameter $Q_{\rm imp}=1$,  for accretion rates of  $10^{-8} M_\odot$ yr$^{-1}$ (green, dotted), $10^{-9} M_\odot$ yr$^{-1}$ (blue, dashed),  and $10^{-10} M_\odot$ yr$^{-1}$ (red, solid).  Right: same, but for a star with a superfluid core.   \label{fig_dT_over_T_v_r_alpha}}  
\end{figure*}

We begin by setting the internal magnetic field strength to zero ($B=0$), so that all asymmetry comes from the non-uniform surface temperature.  Results for $\alpha = 10^{-2}$ are shown in Figure \ref{fig_dT_over_T_v_r_alpha} for stars with normal (left hand plot) and superfluid cores (right hand plot), for $Q_{\rm imp}$ and three different accretion rates.

Extracting the temperature perturbation values at our chosen indicative density of $\rho = 10^{12}$ g cm$^{-3}$, from the left hand plot of Figure \ref{fig_dT_over_T_v_r_alpha}, for normal cores, we obtain
\be
\label{eq:deltaT_over_T_normal_alpha}
\frac{\delta T}{T} \approx 1 \times  10^{-2}   \alpha \Rightarrow \epsilon \approx 5 \times 10^{-8}  \alpha ,
\ee
where we have used equation (\ref{eq:epsilon_T_UCB}) to obtain the second equation.

We can proceed similarly for superfluid cores.  From the right hand plot of Figure  \ref{fig_dT_over_T_v_r_alpha} we obtain
\be
\label{eq:deltaT_over_T_SF_alpha}
\frac{\delta T}{T} \approx 6 \times 10^{-3} \alpha \Rightarrow \epsilon \approx 3 \times 10^{-8} \alpha .
\ee
Clearly, we require asymmetries at the level of  a few tens of percent in $\alpha$ in order to reach the $\epsilon \sim 10^{-8}$ level required for the corresponding gravitational wave emission to provide the necessary braking torque.  Note that the outer edge of our computational domain is located at the density $\rho_{\rm OB} = 1.49 \times 10^{9}$ g cm$^{-3}$, not at the actual surface.  Moving the outer boundary condition closer to the true surface will only serve to make the  actual temperature asymmetries in the deep crust even smaller than those computed here.

These results scale linearly in $\alpha$, while the results of Section  \ref{sect:perturbed_magnetic} are linear in $B$.  It follows we can immediately obtain solutions for non-zero $(B, \alpha)$ through a simple linear combination of the separate $(B, \alpha = 0)$ and $(B=0, \alpha)$ solutions.  

\begin{figure}
\begin{center}
\begin{minipage}[c]{\linewidth}
\includegraphics[width=\textwidth]{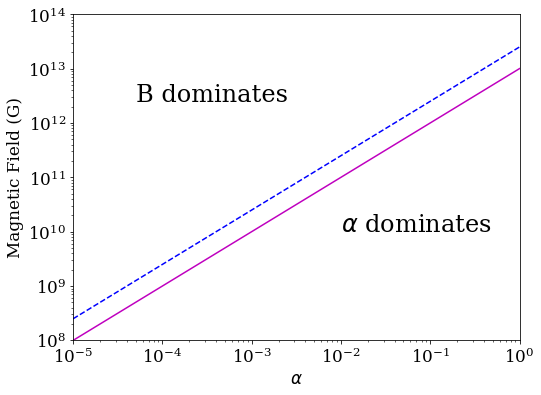}
\caption{Phase space plot  showing which of our two temperature asymmetry mechanisms dominates, {\it viz}  asymmetries due to the internal magnetic field $B$ affecting the thermal conductivity, versus asymmetries due to a non-spherical surface temperature, as parameterised by the accretion asymmertry $\alpha$.  The dividing curve is solid magenta line for a star with a normal core, and dashed blue line for a star with a superfluid core.  We have set an accretion rate $\dot M = 10^{-9} M_\odot$ yr$^{-1}$, and an impurity parameter $Q_{\rm imp} = 1$.  \label{fig_phase_space} }
\end{minipage}
\end{center}
\end{figure}

By combining equations (\ref{eq:deltaT_over_T_normal_B}) and (\ref{eq:deltaT_over_T_normal_alpha}) we obtain the linear relation between $B$ and $\alpha$ such that the two perturbation mechanisms are of equal magnitude for the normal core case.  Specialising to an accretion rate of $\dot M = 10^{-9} M_\odot$ yr$^{-1}$, and $Q_{\rm imp} =1$, we have
\be
B \approx (10^{13} {\, \rm G \,})   \alpha .
\ee
The corresponding result for the superfluid core case is
\be
B \approx (3 \times 10^{13} {\, \rm G \,})   \alpha .
\ee
The corresponding phase space plots are given in Figure \ref{fig_phase_space}.

\section{Possible extensions and improvements}
\label{sect:improvements}

Before summarising and discussing our results in Section \ref{sect:summary}, we first outline a number of areas where, in future work, our model might be made more accurate. 

\emph{Fully self-consistent solutions.}  In our analysis, we computed the temperature perturbation $\delta T$ self-consistently.  However, we then estimated the corresponding ellipticity $\epsilon$ by means of a simple fitting formula to the results of UCB, using the value of the temperature perturbation at a specific density ($\rho = 10^{12}$ g cm$^{3}$) rather than the full function $\delta T(r)$.  A more rigorous treatment would be to use the temperature perturbation computed here as a source term to compute the elastic response, along the lines of the calculations performed in Section 4 of UCB.

\emph{Enlargement of the computational domain.}  Our computational domain extended only over the density range of the \citet{HandZ90a, HandZ90b}  equation of state.  This could be extended to higher densities to move the inner boundary closer to, or at, the centre of the star.  It could be extended to lower densities to move the outer boundary closer to the true stellar surface.  This would make more accurate our association of the outer boundary temperature with the actual, observed, surface temperature.  

\emph{Inclusion of other sources of asymmetry.}  We have allowed for two sources of magnetically-induced asymmetry in our modelling.  In Section  \ref{sect:perturbed_magnetic} we allowed for anisotropy in the thermal conductivity tensor, induced by the internal magnetic field.  In Section \ref{sect:perturbed_surface} we allowed for a surface temperature asymmetry, as might be caused by channeling of the accretion flow by an external magnetic field.  But other sources of asymmetry can be envisaged, possibly also connected with the magnetic field.  These include the local nuclear heating rate and the composition, as discussed in UCB.  Another possibility is the neutrino cooling rate.  However, unlike the case of anisotropic thermal conduction considered here, the precise mechanism by which the magnetic field would induce the asymmetry is not clear.  

\emph{Allowance for superconductivity.}  In our analysis, we implemented a simple prescription to describe the effects of superfluidity.  However, LMXB cores are almost certainly also superconducting.  An analysis taking this into account would be desirable, although the modelling of superconducting magnetic equilibria is still  in its early stages, see e.g.\ \citet{graber_etal_15}.

\emph{Other crustal EoS.}  In building our crust, we used the equation of state of  \citet{HandZ90a, HandZ90b}, which assumes that the ashes of the initial H/He burning are iron $(Z= 26, A=56)$.    The more recent analyses of \citet{HandZ03}  have allowed for higher atomic number ashes,  $(Z= 46 A=106)$, while other works have refined the input nuclear physics \citep{HandZ03, fantina_etal_18}.  Note, however, that the total heat relased per nucleon is probably not very different from the $1.33$ MeV used here.  For instance, \citet{fantina_etal_18} find $1.5$--$1.7$ MeV per nucleon.

\emph{Heat transport by phonons.} In our calculations we took the standard approach of considering heat transport only by electrons.  However, it has been shown by \citet{aguilera_etal_09} that, at the densities and temperatures of interest here, and for magnetic field strengths in excess of $10^{13}$ G, the heat conduction orthogonal to the magnetic field lines can be significantly enhanced by conduction due to superfluid phonons.  Such fields are almost certainly too large to be relevant for LMXBs, but given the strong fields we require to produce significant mountains, and given our ignorance of the strength of the internal magnetic field, inclusion of such a heat transport mechanism may be of interest.  It would presumably act to smooth out the magnetic field-induced asymmetries, reducing the associated mass quadrupoles.

\emph{Fully relativistic calculations.}. As described in Section \ref{sect:hydro}, we use general relativity to compute the hydrostatic structure of the core of our star, but then use Newtonian gravity to compute the  structure of the crust.  The errors made in the hydrostatic structure are modest, as discussed in Section \ref{sect:hydro}.  There will also be errors made in neglecting general relativity when computing the thermal structure.  Given the compactness $M/R \sim 0.2$ of a typical neutron star, one may expect fractional errors at a similar level in the Newtonian thermal calculation.  It would be interesting to carry out a fully relativistic calculation to test this expectation.

\section{Summary and discussion}
\label{sect:summary}

Our main aim in this paper has been to calculate the level of temperature asymmetry generated by an internal magnetic field in an accreting LMXB.  We assumed weak magnetic fields, so that a perturbative approach was possible, and used a fit to the results of UCB that enabled us to connect a given temperature asymmetry $\delta T / T$ to a gravitational wave ellipticity $\epsilon$.  Simple formulae giving $\epsilon$ as a linear function of $B$ are given in equations (\ref{eq:deltaT_over_T_normal_B}) and  (\ref{eq:deltaT_over_T_SF_B}), for normal and superfluid cores respectively, where the accretion-induced heat release of \citet{HandZ90a} are used, corresponding to a release of $1.33$ MeV per accreted nucleon.    If instead one allows for an additional $6$ MeV of heat release at low densities, as is indicated by recently modelling of transiently accretion LMXBs (REF), one instead obtains equations (\ref{eq:deltaT_over_T_normal_B}) and  (\ref{eq:deltaT_over_T_SF_B}).   Taking as representative the superfluid core, non shallow heating case, we found (equation (\ref{eq:deltaT_over_T_SF_B}) above):
\be
\label{eq:deltaT_over_T_SF_B_repeat}
\frac{\delta T}{T} \sim 2 \times 10^{-7} B_9  \Rightarrow \epsilon \approx 1 \times 10^{-12} B_9 .
\ee 
This indicates that internal field strengths of order $10^{13}$ G are required to generate the level of ellipticity needed for gravitational wave emission to halt spin-up.  This is four orders of magnitude stronger than the level of field ($B \sim 10^9$ G) normally expected to exist in an LMXB.  The results for the other cases described above (normal/superfluid core, shallow crustal heating on/off) lead to somewhat different numbers, but the general result remains the same: a very large internal field is required.  

Note that such high field strengths take us out of domain of validity of the perturbative  regime that we used (see equation \ref{eq:omega_tau}), so should be viewed with caution.   It would be interesting to repeat this analysis without making the weak-field approximation, along the linear of the numerical computations that have been carried out for (non-accreting) magnetars, see e.g.\ \citet{Aguilera2008}.  Without such computations to hand, we can not anticipate how the full result will differ from our perturbative one at high magnetic field strengths.

As described in Section \ref{sect:number}, before carrying out detailed calculations, we estimated the magnetisation parameter, $\omega \tau$, in an attempt to estimate the likely level of temperature asymmetry.  We found $\omega \tau \sim 10^{-4} B_9$; see equation (\ref{eq:omega_tau}) for the full result.  This is about three orders of magnitude larger than the temperature asymmetries computed via detailed calculations.  We speculate that this discrepancy may be due to the magnetisation parameter only capturing the effect of the magnetic field on the heat conduction.  There is also a large heat flux due to neutrinos, which is not (in our model) affected by the magnetic field.  Note that, in their work, UCB considered  compositional and nuclear burning  asymmetries, which they parameterised in terms of dimensionless numbers $f_{\rm comp}$ and $f_{\rm nuc}$.  They too found that the temperature asymmetries were significantly smaller than their dimensionless parameterisations would suggest, although not by the large margin that we find here for magnetic conductivity asymmetries.

It is interesting to compare the ellipticities computed here with those more normally associated with magnetised stars, i.e.\ with the mountains  that are created \emph{directly} by the magnetic field, via Lorentz forces.  For a star with a non-superconducting core, one can estimate the ellipticity by taking the ratio of the magnetostatic energy to the gravitational binding energy (see e.g.\ \citet{Cutler2002} and references therein):
\be
\epsilon \approx \frac{B^2 R^4}{G M^2} \approx 2 \times 10^{-18}  B_9^2 , 
\ee
a result that is confirmed by detailed numerical simulation \citep{LandJ_09}.  Probably more relevant for the LMXBs is the corresponding result for a star with a superconducting core, where one factor of $B$ is replaced by the `first critical field' $H_{c1} \approx 10^{15}$ G \citep{Cutler2002}:
\be
\label{eq:epsilon_supercon}
\epsilon \approx  \frac{B H_{\rm c} R^4}{G M^2} \approx 2 \times 10^{-12}  B_9 .
\ee
Comparing equations (\ref{eq:deltaT_over_T_SF_B_repeat})  and  (\ref{eq:epsilon_supercon}), it is interesting to note that the two results are rather similar, with the same linear scaling in $B$ and with the same order of magnitude.  This is, of course, a coincidence, as they are based on completely different physical mechanisms. 

As discussed in UCB,  temperature asymmetries will give rise to modulation in the surface flux, potentially leading to an observable  modulation in the X-ray flux from the rotating star.  As noted in UCB, for steadily accreting systems, the modulation will be small compared to the accretion flux ($\sim 200$ MeV per nucleon), and therefore probably unobservable.  For a very large internal field of $B \sim 10^{12}$ G, we find a fractional modulation of $\sim 10^{-3}$.  However, if accretion were to stop, the relevant comparison is with the spherical thermal flux through the surface of the unmagnetised star  ($F$, in our notation), which is about two orders of magnitude  smaller, giving rise to a correspondingly larger fractional modulation (see discussion in UCB and \citet{Haskell2015}).  However, in this case, a proper treatment of the temperature gradient in the outer layers would be needed to make proper contact with observations.  Nevertheless, new and upcoming X-ray instruments, such as NICER, may be able to constrain such variations.

As noted in Section \ref{sect:intro}, \citet{singh_etal_19} recently considered a very different mechanism for generating temperature asymmetries in magnetised LMXBs.   They considered the displacement in capture layers by the direct magnetic deformation from a magnetically-confined mound, and argued that this itself induced a temperature asymmetry.  In the absence of shallow crustal heating, they found asymmetries too small to allow for significant gravitational wave emission.  However, in the case of shallow crustal heating, they found, when extrapolating their results to densities greater than those captured in their computational domain, asymmetries large enough to make gravitational wave emission significant in both steadily accreting systems, and also in the transiently accreting pulsar PSR J1023+0038, whose possible relevance in terms of such gravitational wave emission was already pointed out in \citet{haskell_patruno_17}.  It will be interesting to see if this result persist in a fully self-consistent treatment, carried out over the full density range.

We also examined how asymmetries in the surface temperature would induce internal temperature asymmetries, and therefore generate mountains.  Such surface temperature asymmetries could be caused by asymmetric accretion, perhaps funnelled by the external dipolar magnetic field.  For an accretion asymmetry of fractional size $\alpha$, we found, for stars with superconducting cores and no shallow crustal heating,  
\be
\label{eq:deltaT_over_T_SF_alpha_repeat}
\frac{\delta T}{T} \approx 6 \times 10^{-3} \alpha \Rightarrow \epsilon \approx 3 \times 10^{-8} \alpha .
\ee
Clearly, a very large surface temperature asymmetry is required to produce the ellipticity required. 

To sum up, our main finding is that for our model of anisotropic thermal conduction, to produce mountains at an interesting level for gravitational wave emission, a very strong internal magnetic field is required, of order $10^{13}$ G.  This has to be taken with caution, as this field strength is large enough to in fact push us out of the perturbative regime in which we work.   This is about four orders of magnitude  larger than the magnetic fields normally associated with LMXBs.   Note, however, that the most relevant field component for our model is the internal toroidal quadrupolar component.  We do not actually have any direct observational constraints on this, as accretion spin-up and pulsar spin-down is expected to be dominated by the external poloidal dipolar component.  The possibility of the internal fields being larger than the external ones certainly cannot be discounted, as magnetic fields can be buried by the accretion process itself \citep{mukherjee_17}.  Future gravitational wave observations will provide a probe of this scenario.

\section*{Acknowledgements}

The authors acknowledge support from STFC via grant number ST/R00045X/1, and travel support from the PHAROS COST Action (CA16214).   They also acknowledge useful discussions with Nicloas Chamel, Andrew Cumming, Anthea Fantina, Ian Hawke, Thomas Hutchins and Neha Singh.

\bibliographystyle{mnras}


\bibliography{library}

\begin{thebibliography}{}
\makeatletter
\relax
\def\mn@urlcharsother{\let\do\@makeother \do\$\do\&\do\#\do\^\do\_\do\%\do\~}
\def\mn@doi{\begingroup\mn@urlcharsother \@ifnextchar [ {\mn@doi@}
  {\mn@doi@[]}}
\def\mn@doi@[#1]#2{\def\@tempa{#1}\ifx\@tempa\@empty \href
  {http://dx.doi.org/#2} {doi:#2}\else \href {http://dx.doi.org/#2} {#1}\fi
  \endgroup}
\def\mn@eprint#1#2{\mn@eprint@#1:#2::\@nil}
\def\mn@eprint@arXiv#1{\href {http://arxiv.org/abs/#1} {{\tt arXiv:#1}}}
\def\mn@eprint@dblp#1{\href {http://dblp.uni-trier.de/rec/bibtex/#1.xml}
  {dblp:#1}}
\def\mn@eprint@#1:#2:#3:#4\@nil{\def\@tempa {#1}\def\@tempb {#2}\def\@tempc
  {#3}\ifx \@tempc \@empty \let \@tempc \@tempb \let \@tempb \@tempa \fi \ifx
  \@tempb \@empty \def\@tempb {arXiv}\fi \@ifundefined
  {mn@eprint@\@tempb}{\@tempb:\@tempc}{\expandafter \expandafter \csname
  mn@eprint@\@tempb\endcsname \expandafter{\@tempc}}}

\bibitem[\protect\citeauthoryear{Abbott et~al.}{Abbott
  et~al.}{2017}]{LVC_GWs_and_GRBs_17}
Abbott B.,  et~al., 2017, \mn@doi [\apjl] {10.3847/2041-8213/aa920c}, \href
  {https://ui.adsabs.harvard.edu/abs/2017ApJ...848L..13A} {848, L13}

\bibitem[\protect\citeauthoryear{Abbott et~al.}{Abbott
  et~al.}{2018a}]{LVC_catalog_18}
Abbott B.,  et~al., 2018a, arXiv e-prints, \href
  {https://ui.adsabs.harvard.edu/abs/2018arXiv181112907T} {p. arXiv:1811.12907}

\bibitem[\protect\citeauthoryear{Abbott et~al.}{Abbott
  et~al.}{2018b}]{LVC_BBH_populations_18}
Abbott B.,  et~al., 2018b, arXiv e-prints, \href
  {https://ui.adsabs.harvard.edu/abs/2018arXiv181112940T} {p. arXiv:1811.12940}

\bibitem[\protect\citeauthoryear{Abbott et~al.}{Abbott
  et~al.}{2018c}]{LVC_EoS_18}
Abbott B.,  et~al., 2018c, \mn@doi [\prl] {10.1103/PhysRevLett.121.161101},
  \href {https://ui.adsabs.harvard.edu/abs/2018PhRvL.121p1101A} {121, 161101}

\bibitem[\protect\citeauthoryear{Abbott et~al.}{Abbott
  et~al.}{2019a}]{LVC_SCo-X1-O2_19}
Abbott B.,  et~al., 2019a, arXiv e-prints, \href
  {https://ui.adsabs.harvard.edu/abs/2019arXiv190612040T} {p. arXiv:1906.12040}

\bibitem[\protect\citeauthoryear{Abbott et~al.}{Abbott
  et~al.}{2019b}]{LVC_allsky_isolated_O2_19}
Abbott B.,  et~al., 2019b, \mn@doi [\prd] {10.1103/PhysRevD.100.024004}, \href
  {https://ui.adsabs.harvard.edu/abs/2019PhRvD.100b4004A} {100, 024004}

\bibitem[\protect\citeauthoryear{Abbott et~al.}{Abbott
  et~al.}{2019c}]{LVC_testing_GR_19}
Abbott B.,  et~al., 2019c, \mn@doi [\prl] {10.1103/PhysRevLett.123.011102},
  \href {https://ui.adsabs.harvard.edu/abs/2019PhRvL.123a1102A} {123, 011102}

\bibitem[\protect\citeauthoryear{Abbott et~al.}{Abbott
  et~al.}{2019d}]{LVC_known_pulsars_O1-O2_19}
Abbott B.,  et~al., 2019d, \mn@doi [\apj] {10.3847/1538-4357/ab20cb}, \href
  {https://ui.adsabs.harvard.edu/abs/2019ApJ...879...10A} {879, 10}

\bibitem[\protect\citeauthoryear{Aguilera, Pons  \& Miralles}{Aguilera
  et~al.}{2008}]{Aguilera2008}
Aguilera D.~N.,  Pons J.~A.,   Miralles J.~A.,  2008, \mn@doi [Astronomy and
  Astrophysics] {10.1051/0004-6361:20078786}, 486, 255

\bibitem[\protect\citeauthoryear{{Aguilera}, {Cirigliano}, {Pons}, {Reddy}  \&
  {Sharma}}{{Aguilera} et~al.}{2009}]{aguilera_etal_09}
{Aguilera} D.~N.,  {Cirigliano} V.,  {Pons} J.~A.,  {Reddy} S.,   {Sharma} R.,
  2009, \mn@doi [\prl] {10.1103/PhysRevLett.102.091101}, \href
  {https://ui.adsabs.harvard.edu/abs/2009PhRvL.102i1101A} {102, 091101}

\bibitem[\protect\citeauthoryear{Bildsten}{Bildsten}{1998}]{Bildsten1998}
Bildsten L.,  1998, \mn@doi [The Astrophysical Journal] {10.1086/311440}, 501,
  L89

\bibitem[\protect\citeauthoryear{{Brown}}{{Brown}}{2000}]{brown_00}
{Brown} E.~F.,  2000, \mn@doi [\apj] {10.1086/308487}, \href
  {https://ui.adsabs.harvard.edu/abs/2000ApJ...531..988B} {531, 988}

\bibitem[\protect\citeauthoryear{Brown \& Cumming}{Brown \&
  Cumming}{2009}]{Brown2009}
Brown E.~F.,  Cumming A.,  2009, \mn@doi [The Astrophysical Journal]
  {10.1088/0004-637x/698/2/1020}, 698, 1020

\bibitem[\protect\citeauthoryear{Cutler}{Cutler}{2002}]{Cutler2002}
Cutler C.,  2002, \mn@doi [Physical Review D] {10.1103/physrevd.66.084025}, 66

\bibitem[\protect\citeauthoryear{Deibel, Cumming, Brown  \& Page}{Deibel
  et~al.}{2015}]{Deibel2015}
Deibel A.,  Cumming A.,  Brown E.~F.,   Page D.,  2015, \mn@doi [The
  Astrophysical Journal] {10.1088/2041-8205/809/2/l31}, 809, L31

\bibitem[\protect\citeauthoryear{{Douchin} \& {Haensel}}{{Douchin} \&
  {Haensel}}{2001}]{douchin_haensel_01}
{Douchin} F.,  {Haensel} P.,  2001, \mn@doi [\aap]
  {10.1051/0004-6361:20011402}, \href
  {https://ui.adsabs.harvard.edu/abs/2001A&A...380..151D} {380, 151}

\bibitem[\protect\citeauthoryear{{Fantina}, {Zdunik}, {Chamel}, {Pearson},
  {Haensel}  \& {Goriely}}{{Fantina} et~al.}{2018}]{fantina_etal_18}
{Fantina} A.~F.,  {Zdunik} J.~L.,  {Chamel} N.,  {Pearson} J.~M.,  {Haensel}
  P.,   {Goriely} S.,  2018, \mn@doi [\aap] {10.1051/0004-6361/201833605},
  \href {https://ui.adsabs.harvard.edu/abs/2018A&A...620A.105F} {620, A105}

\bibitem[\protect\citeauthoryear{{Glampedakis} \& {Gualtieri}}{{Glampedakis} \&
  {Gualtieri}}{2018}]{glampedakis_gualtieri_18}
{Glampedakis} K.,  {Gualtieri} L.,  2018, in {Rezzolla} L.,  {Pizzochero} P.,
  {Jones} D.~I.,  {Rea} N.,   {Vida{\~n}a} I.,  eds,  Astrophysics and Space
  Science Library Vol. 457, Astrophysics and Space Science Library. p.~673
  (\mn@eprint {arXiv} {1709.07049}), \mn@doi{10.1007/978-3-319-97616-7_12}

\bibitem[\protect\citeauthoryear{{Graber}, {Andersson}, {Glampedakis}  \& {Land
  er}}{{Graber} et~al.}{2015}]{graber_etal_15}
{Graber} V.,  {Andersson} N.,  {Glampedakis} K.,   {Land er} S.~K.,  2015,
  \mn@doi [\mnras] {10.1093/mnras/stv1648}, \href
  {https://ui.adsabs.harvard.edu/abs/2015MNRAS.453..671G} {453, 671}

\bibitem[\protect\citeauthoryear{Haensel \& Zdunik}{Haensel \&
  Zdunik}{1990a}]{HandZ90a}
Haensel P.,  Zdunik J.~L.,  1990a, Astronomy {\&} Astrophysics, 227, 431

\bibitem[\protect\citeauthoryear{Haensel \& Zdunik}{Haensel \&
  Zdunik}{1990b}]{HandZ90b}
Haensel P.,  Zdunik J.~L.,  1990b, Astronomy and Astrophysics, 229, 117

\bibitem[\protect\citeauthoryear{{Haensel} \& {Zdunik}}{{Haensel} \&
  {Zdunik}}{2003}]{HandZ03}
{Haensel} P.,  {Zdunik} J.~L.,  2003, \mn@doi [\aap]
  {10.1051/0004-6361:20030708}, \href
  {https://ui.adsabs.harvard.edu/abs/2003A&A...404L..33H} {404, L33}

\bibitem[\protect\citeauthoryear{Haensel, Kaminker  \& Yakovlev}{Haensel
  et~al.}{1996}]{HKY1996}
Haensel P.,  Kaminker A.~D.,   Yakovlev D.~G.,  1996, Astronomy and
  Astrophysics, 314, 328

\bibitem[\protect\citeauthoryear{{Haskell} \& {Patruno}}{{Haskell} \&
  {Patruno}}{2017}]{haskell_patruno_17}
{Haskell} B.,  {Patruno} A.,  2017, \mn@doi [\prl]
  {10.1103/PhysRevLett.119.161103}, \href
  {https://ui.adsabs.harvard.edu/abs/2017PhRvL.119p1103H} {119, 161103}

\bibitem[\protect\citeauthoryear{Haskell, Priymak, Patruno, Oppenoorth, Melatos
   \& Lasky}{Haskell et~al.}{2015}]{Haskell2015}
Haskell B.,  Priymak M.,  Patruno A.,  Oppenoorth M.,  Melatos A.,   Lasky
  P.~D.,  2015, \mn@doi [Monthly Notices of the Royal Astronomical Society]
  {10.1093/mnras/stv726}, 450, 2393

\bibitem[\protect\citeauthoryear{{Lander} \& {Jones}}{{Lander} \&
  {Jones}}{2009}]{LandJ_09}
{Lander} S.~K.,  {Jones} D.~I.,  2009, \mn@doi [\mnras]
  {10.1111/j.1365-2966.2009.14667.x}, \href
  {https://ui.adsabs.harvard.edu/abs/2009MNRAS.395.2162L} {395, 2162}

\bibitem[\protect\citeauthoryear{{Mukherjee}}{{Mukherjee}}{2017}]{mukherjee_17}
{Mukherjee} D.,  2017, \mn@doi [Journal of Astrophysics and Astronomy]
  {10.1007/s12036-017-9465-6}, \href
  {https://ui.adsabs.harvard.edu/abs/2017JApA...38...48M} {38, 48}

\bibitem[\protect\citeauthoryear{Pons \& Geppert}{Pons \&
  Geppert}{2007}]{Pons07}
Pons J.~A.,  Geppert U.,  2007, \mn@doi [Astron.Astrophys.]
  {10.1051/0004-6361:20077456}, 470

\bibitem[\protect\citeauthoryear{Schatz, Bildsten, Cumming  \& Wiescher}{Schatz
  et~al.}{1999}]{Schatz_1999}
Schatz H.,  Bildsten L.,  Cumming A.,   Wiescher M.,  1999, \mn@doi [The
  Astrophysical Journal] {10.1086/307837}, 524, 1014

\bibitem[\protect\citeauthoryear{Shapiro \& Teukolsky}{Shapiro \&
  Teukolsky}{1983}]{SandT}
Shapiro S.~L.,  Teukolsky S.~A.,  1983, Black Holes, White Dwarfs, and Neutron
  Stars.
Wiley-Verlag, \mn@doi{10.1002/9783527617661}

\bibitem[\protect\citeauthoryear{{Singh}, {Haskell}, {Mukherjee}  \&
  {Bulik}}{{Singh} et~al.}{2019}]{singh_etal_19}
{Singh} N.,  {Haskell} B.,  {Mukherjee} D.,   {Bulik} T.,  2019, arXiv
  e-prints, \href {https://ui.adsabs.harvard.edu/abs/2019arXiv190805038S} {p.
  arXiv:1908.05038}

\bibitem[\protect\citeauthoryear{{Urpin} \& {Yakovlev}}{{Urpin} \&
  {Yakovlev}}{1980}]{yakovlev_urpin_80}
{Urpin} V.~A.,  {Yakovlev} D.~G.,  1980, \sovast, \href
  {https://ui.adsabs.harvard.edu/abs/1980SvA....24..425U} {24, 425}

\bibitem[\protect\citeauthoryear{Ushomirsky, Cutler  \& Bildsten}{Ushomirsky
  et~al.}{2000}]{UCB}
Ushomirsky G.,  Cutler C.,   Bildsten L.,  2000, \mn@doi [Monthly Notices of
  the Royal Astronomical Society] {10.1046/j.1365-8711.2000.03938.x}, 319, 902

\makeatother
\end{thebibliography}


\appendix

\label{lastpage}

\end{document}